\documentclass{iau}
\usepackage{graphicx}
\usepackage{natbib}

\def\mnras{MNRAS}
\def\apj{ApJ}
\def\apjs{ApJS}
\def\araa{ARAA}
\def\apjl{ApJL}
\def\nat{Nature}
\def\aap{A\&A}
\def\aj{AJ}

\title[Modelling the formation of today's massive ellipticals] 
{Modelling the formation of today's massive ellipticals}

\author[Thorsten Naab]   
{Thorsten Naab$^1$} 

\affiliation{$^1$Max-Planck-Institute for Astrophysics,
  Karl-Schwarzschild-Str. 1, 85741 Garching, Germany \\ email: {\tt
    naab@mpa-garching.mpg.de}}

\pubyear{2013}
\volume{295}  
\pagerange{xx--xx}
\jname{The intriguing life of massive galaxies}
\editors{D. Thomas, A. Pasquali \& I. Ferreras, eds.}
\begin{document}

\maketitle

\begin{abstract}
The discovery of a population of massive, compact and quiescent
early-type galaxies has changed the view on plausible formation
scenarios for the present day population of elliptical
galaxies. Traditionally assumed formation histories dominated by
'single events' like early collapse or major mergers appear to be
incomplete and have to be embedded in the context of hierarchical
cosmological models with continuous gas accretion and the merging of
small stellar systems (minor mergers). Once these processes are
consistently taken into account the hierarchical models favor a
two-phase assembly process and are in much
better shape to capture the observed trends. We review some aspects of
recent progress in the field.   
   
\keywords{galaxies: elliptical and lenticular, galaxies: formation, galaxies: evolution}
\end{abstract}

\firstsection 
\section{Introduction}
During the formation and assembly of massive galaxies merging is a
natural process in modern hierarchical cosmological models. It is
expected to play a significant role for the structural and
morphological evolution
(e.g. \citealp{1996MNRAS.283L.117K,1996MNRAS.281..487K,2006MNRAS.366..499D,2006ApJ...648L..21K,2007MNRAS.375....2D,2008MNRAS.384....2G,2009ApJS..182..216K,2010ApJ...715..202H}). In 
the light of these theoretical expectations and direct observations of 
'dry' mergers of gas poor elliptical galaxies up to high redshift
\citep{2005AJ....130.2647V,2005ApJ...627L..25T,2006ApJ...640..241B,2006ApJ...652..270B,2008ApJ...672..177L,2009ApJ...697.1971J,2012ApJ...746..162N,2012ApJ...744...85M}
simulations of idealized collisionless mergers have again received
attention and new studies were triggered. Merger simulations of already
existing spheroidal galaxies have focused in detail on the evolution
of abundance gradients, shapes and kinematics, scaling relations, sizes and dark 
matter fractions  
\citep{1978MNRAS.184..185W,1979MNRAS.189..831W,1997ApJ...481...83M,2005MNRAS.362..184B,2006ApJ...636L..81N,2006MNRAS.369.1081B,2008MNRAS.383...93B,2009A&A...501L...9D,2009ApJ...703.1531N,2012arXiv1202.0971N}. 
If the progenitors were two disk galaxies (which then can include a
gaseous component) the aim was to investigate the morphological
transformation, i.e. the formation of new dynamically hot spheroidal
elliptical galaxies from two dynamically cold progenitor spiral 
galaxies
\citep{1981MNRAS.197..179G,1982ApJ...259..103F,1983MNRAS.205.1009N,1988ApJ...331..699B,1992ARA&A..30..705B,1992ApJ...400..460H}. Apart
from studies of the effect of the merger mass-ratio
\citep{1998giis.conf..275B,1998ApJ...502L.133B,2000MNRAS.316..315B,2003ApJ...597..893N,2004A&A...418L..27B,2005A&A...437...69B,2005MNRAS.357..753G}
the tidal torquing of gas, its inflow to the central regions, the
impact on the stellar
orbits \citep{1996ApJ...471..115B,2006MNRAS.372..839N,2010ApJ...723..818H},
subsequent starbursts
\citep{1994ApJ...431L...9M,1996ApJ...464..641M,2004MNRAS.350..798B,2008A&A...492...31D}
and the potential growth of black holes
\citep{1989Natur.340..687H,2005MNRAS.361..776S,2005Natur.433..604D,2009ApJ...690..802J,2009MNRAS.396L..66Y}
was investigated in numerous studies together with influential studies
on the origin of early-type galaxy scaling relations
\citep{2006ApJ...641...21R,2006MNRAS.370.1445D,2006ApJ...650..791C,2008ApJ...679..156H,2009ApJ...691.1424H,2009ApJS..181..135H,2009ApJS..181..486H,2010MNRAS.406L..55D,2011MNRAS.415.3750M}. 
However, despite the detailed insights on the stellar and gas dynamical
processes in simulated galaxy mergers, the 'binary merger' approach
is limited in scope and seems not to be able to naturally
explain all properties of present day massive elliptical galaxies
\citep{2009ApJ...690.1452N}.  

The most massive elliptical galaxies (or their
progenitors) are considered to start forming their stars at high
redshift ($z\sim 6$, or higher) in a dissipative environment, rapidly
become very massive ($\sim 10^{11}M_{\odot}$) by $z =2$  
\citep{2005MNRAS.363....2K,2006ApJ...648L..21K,2006MNRAS.366..499D,2006ApJ...649L..71K,2007ApJ...658..710N,2009ApJ...699L.178N,2009ApJ...692L...1J,2009ApJ...703..785D,2009MNRAS.395..160K,2010ApJ...725.2312O,2010ApJ...709..218F,2011MNRAS.417..900D,2011ApJ...736...88F,2012ApJ...744...63O}. A
significant fraction of this high redshift population is observed to 
be already quiescent at $z \sim 2$, on 
average 4-5 times more compact (part of this apparent evolution might
driven by selection effects, see e.g. \citealp{2012arXiv1211.1005P}),
and typically a factor of two less massive than their low redshift descendants 
\citep{2005ApJ...626..680D,2005ApJ...631..145V,2005A&A...442..125D,2006ApJ...650...18T,2007MNRAS.374..614L, 
2007ApJ...671..285T,2008ApJ...687L..61B,2008ApJ...677L...5V,2008ApJ...688...48V,2008A&A...482...21C,2008ApJ...688..770F,2009ApJ...695..101D,2009ApJ...696L..43C,2009ApJ...697.1290B, 
2010ApJ...709.1018V,2011ApJ...736L...9V,2012ApJ...745..179W}. It
is reasonable to assume that the high-redshift population forms the cores of at
least some, if not all, present day massive ellipticals. This rapid
structural evolution is supposed to happen in an   
inside-out fashion, mainly by adding stellar mass to the outer parts
of the galaxies over time, however, without the formation of a
significant fraction of new stars
\citep{2009MNRAS.398..898H,2010ApJ...709.1018V,2012ApJ...749..121S,2012MNRAS.422.3107S}. In
this respect the growth of massive quiescent high-redshift galaxies is
markedly different to the star formation  driven inside-out growth of disk
galaxies

The implications of these observational findings for the formation and
evolution of massive elliptical galaxies are many-fold. They
are unlikely to have formed by an initial 'monolithic collapse'
followed by passive evolution as their present day counterparts would
be too small and too red
\citep{2008ApJ...677L...5V,2008ApJ...682..896K,2009ApJ...697.1290B,2012MNRAS.tmp.2790F}. In addition the
evolution of these system cannot be explained by just a single 'binary
merger of disk galaxies'. The compact high-redshift systems might
have formed in such a process \citep{2010ApJ...722.1666W,2011ApJ...730....4B}, if it were
gas-rich, but the subsequent structural evolution requires additional
processes which are not driven by the formation of new
stars. Observational results that almost none of these massive compact
galaxies were able to survive to the present day
\citep{2009ApJ...692L.118T,2010ApJ...720..723T} indicate that a general and common
physical mechanism must be 
at work. Spectacular events alone, like major early-type galaxy mergers, might be too
rare.

\section{Minor mergers vs. major mergers}

Minor merges, however, are expected to happen frequently in the
lifetime of a massive galaxy and have received particular attention as
they provide a natural way to increase the size of a galaxy. With only
a few assumptions the virial theorem provides a simple estimate of how
a one-component system evolves during major and minor mergers
\citep{2000MNRAS.319..168C,2009ApJ...699L.178N,2009ApJ...697.1290B}. Following
\citet{2009ApJ...699L.178N} we assume that a compact initial stellar system has formed
(e.g. involving gas dissipation) with  a total energy $E_i$, a mass
$M_i$, a gravitational radius $r_{g,i}$, and the mean  square speed of
the stars is $\langle v_i^2 \rangle $. According to the virial theorem
\citep{2008gady.book.....B} the total energy of the system is  

\begin{eqnarray} 
E_i & = &  K_i+W_i = -K_i = \frac{1}{2} W_i \nonumber \\
    & = & -\frac{1}{2} M_i \langle v_i^2 \rangle = -\frac{1}{2}
\frac{GM_i^2}{r_{g,i}}. \nonumber
\end{eqnarray}

This system then merges (on zero energy orbits) with other systems of
a total energy $E_a$, total mass  $M_a$, gravitational radii
$r_{a}$ and mean square speeds averaging $\langle  v_a^2\rangle$. The
fractional mass increase from all the merged galaxies is $\eta =
M_a/M_i$ and the total kinetic energy of the material is $K_a=(1/2)
M_a \langle v_a^2\rangle$, further defining  $\epsilon = \langle v_a^2
\rangle/\langle v_i^2\rangle $. Under the assumption of energy
conservation (results from \citep{2006A&A...445..403K}
indicate that most halos merge on parabolic orbits) the ratio of
initial to final mean square speeds, gravitational radii and densities
can be then written as \citep{2009ApJ...699L.178N}

\begin{eqnarray}
\frac{\langle v_f^2\rangle }{\langle v_i^2\rangle } =
\frac{(1+\eta\epsilon)}{1+\eta} , 
\frac{r_{g,f}}{r_{g,i}} = \frac{(1+\eta)^2}{(1+\eta\epsilon)},
\frac{\rho_f}{\rho_i} = \frac{(1+\eta
  \epsilon)^3}{(1+\eta)^5}. \nonumber
\end{eqnarray}

For mergers of two identical systems, $\eta = 1$, the mean square
speed would remain unchanged, the size increases by a factor of two
and the densities drop by a factor of four. In the limit that the mass
is accreted in the form of very weakly bound stellar systems with
$\langle v_a^2\rangle  << \langle v_i^2\rangle $ or $\epsilon << 1$,
the mean square speed is reduced by a factor two, the size increases
by a factor four and the density drops by a factor of 32. These
estimates are, however, idealized assuming one-component systems, no
violent relaxation and zero-energy orbits with fixed angular
momentum. 

\begin{figure}[t]
\begin{center}
 \includegraphics[width=2.6in]{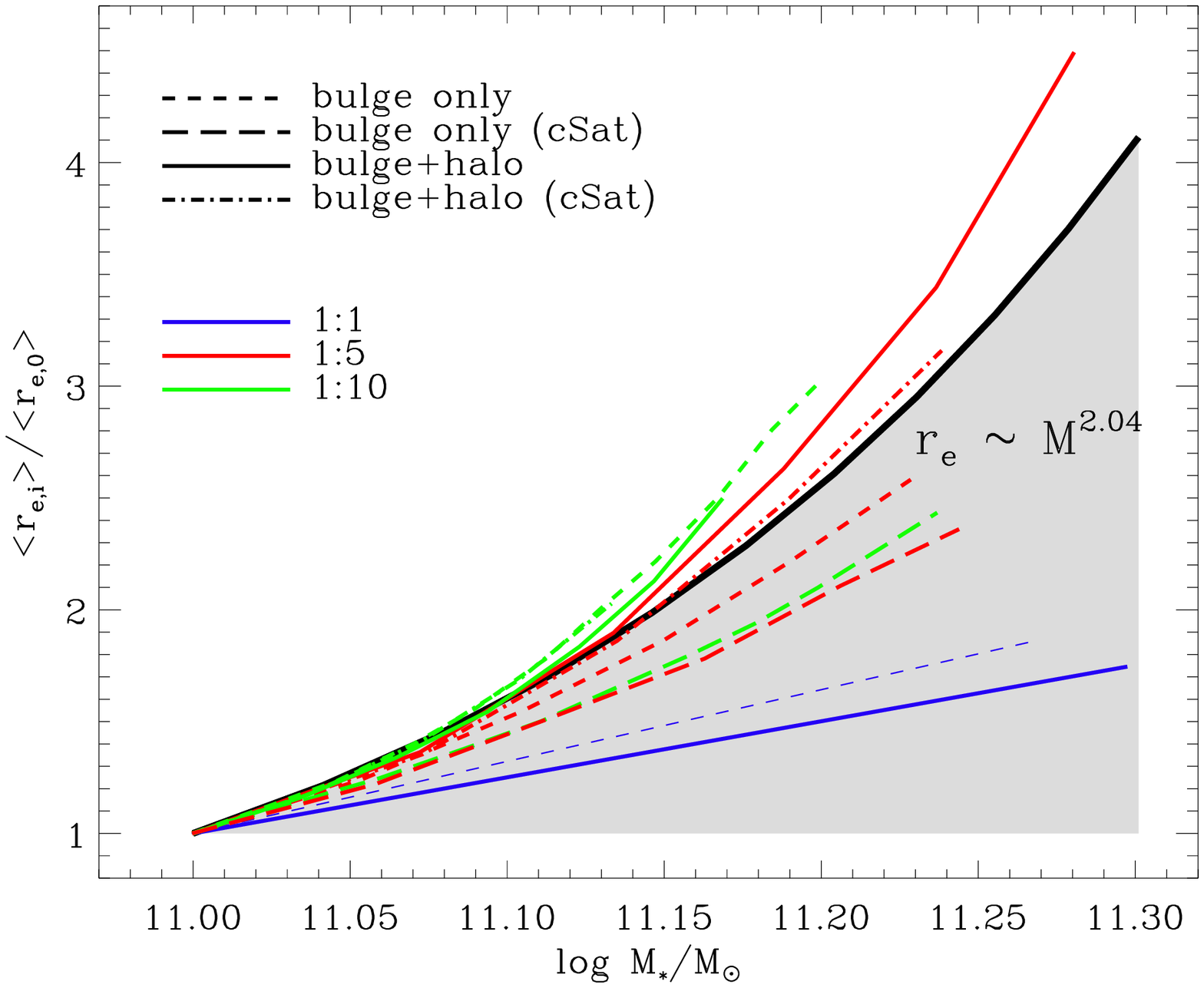} 
 \includegraphics[width=2.6in]{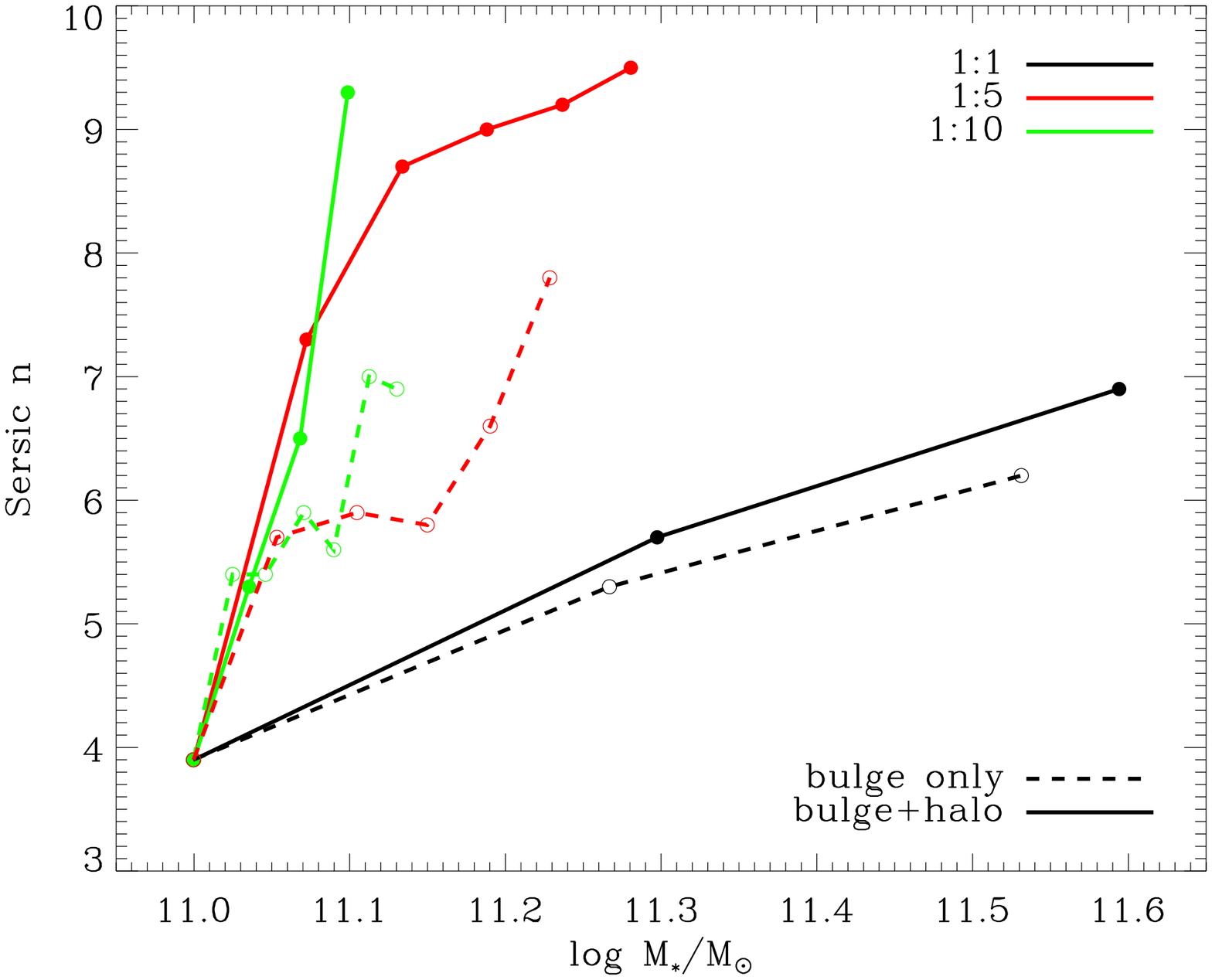} 

 \caption{{\it Left:} Simulated size evolution as a function of bound
   stellar mass for mergers with mass-ratios 1:1 (blue), 5:1 (red),
   and 10:1 (green). The observationally expected relation is
   indicated by the black line \citep{2010ApJ...709.1018V}. The
   presence of dark matter significantly boosts the size evolution of
   5:1 and 10:1 mergers. {\it Right:} This also leads to a significantly
   stronger evolution of the Sersic index (figures taken from
   \citet{2012arXiv1206.5004H})  
   \label{fig1}}
\end{center}
\end{figure}

\citet{2012MNRAS.425.3119H} have recently re-investigated in
detail the collisionless dynamics of major and minor mergers of
systems including concentrated stellar spheroidal components embedded
in extended dark matter halos. They present more accurate versions of
the above equations including the effect of escapers and the
interaction of the stellar baryonic with dark matter and describe in
detail how the presence of a massive dark matter halos alter the
evolution of the merging systems. One result of this study was that
both minor and major mergers lead to size growth and an increase of
the dark matter fraction. The physical processes are, however,
different. Violent relaxation in major mergers mixes dark matter to 
the central regions. Escaping, unbound, particles limit the expected
size growth to values below the ones expected from the idealized
equations above. In minor mergers (mass-ratios of 1:5 and 1:10), the
stellar satellites are stripped at large radii where the host
galaxies dominated by dark matter and the stellar effective
radii and the dark matter fractions grow more rapidly than
expected from the simple virial equations (see also
\citealp{2012MNRAS.424..747L}). Due to the addition of stellar
satellite material at large radii \citep{1983MNRAS.204..219V}, the
stellar mass distribution changes 
significantly resulting in a significant increase of the Sersic index
(see Fig. \ref{fig1} and \citealp{2012arXiv1206.5004H}). 
The general results on size evolution are in agreement with similar
studies by e.g. \citet{2012MNRAS.tmp...42O}. However, there is an
ongoing debate of whether the size growth by minor mergers is
sufficient to explain the observed cosmological size evolution of
elliptical galaxies. Whereas \citet{2012MNRAS.tmp...42O} argue that
the size growth by minor mergers alone might be sufficient, studies by 
\citet{2012MNRAS.422.1714N}, \citet{2012MNRAS.422L..62C}, and 
\citet{2012ApJ...746..162N} have combined idealized numerical
simulations embedded in a cosmological context and new observational
constraints. They come to the conclusion that minor mergers might be
able to explain the observed size growth from redshift $z \sim 1$ to
the present. However, at higher redshift minor and major mergers might
not be frequent enough to explain the rapid size evolution observed at
$z \gtrsim 1$ and therefore an additional physical mechanism might be
required.   

\begin{figure}[t]
\begin{center}
 \includegraphics[width=2.5in]{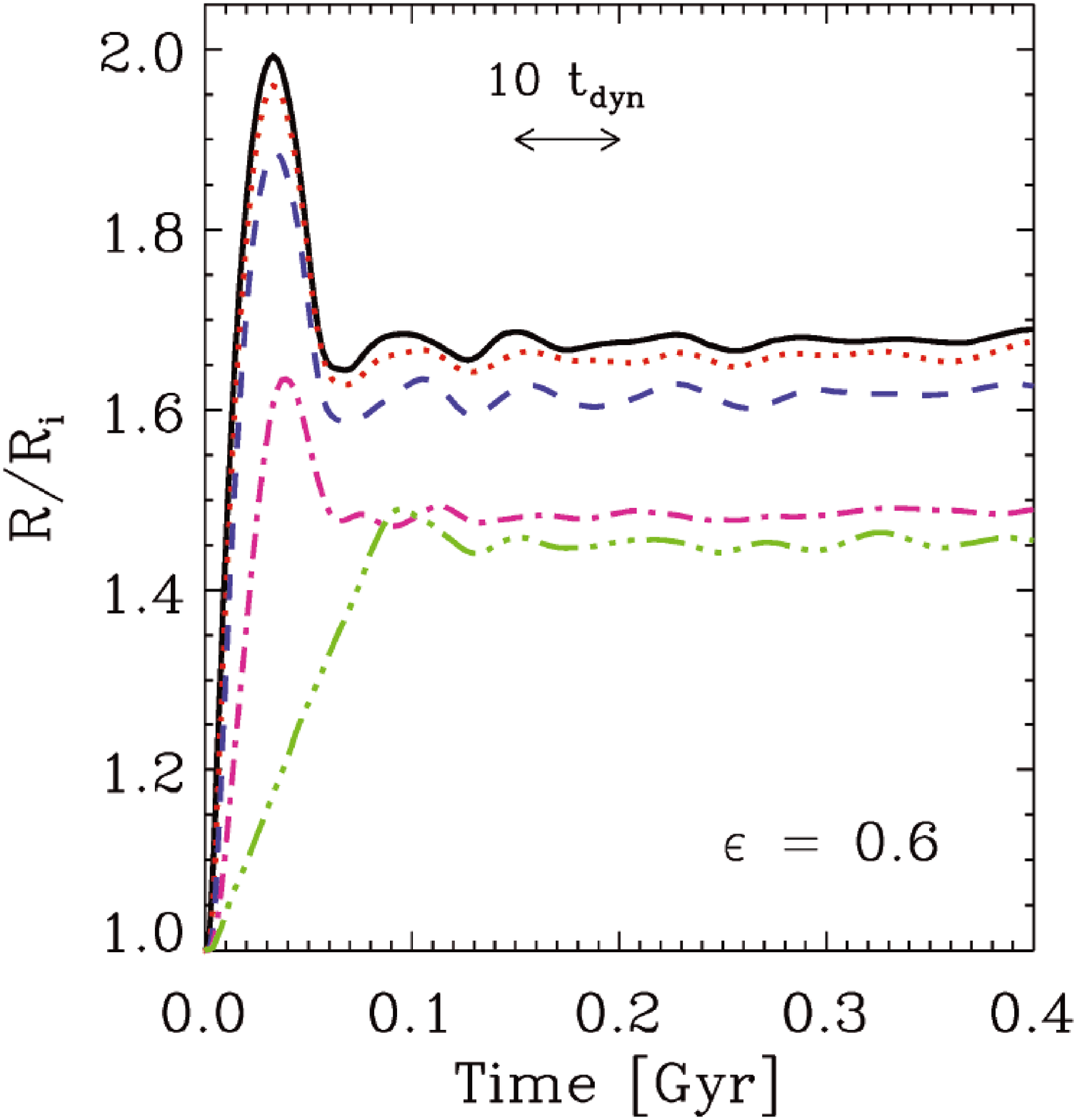} 
 \includegraphics[width=2.7in]{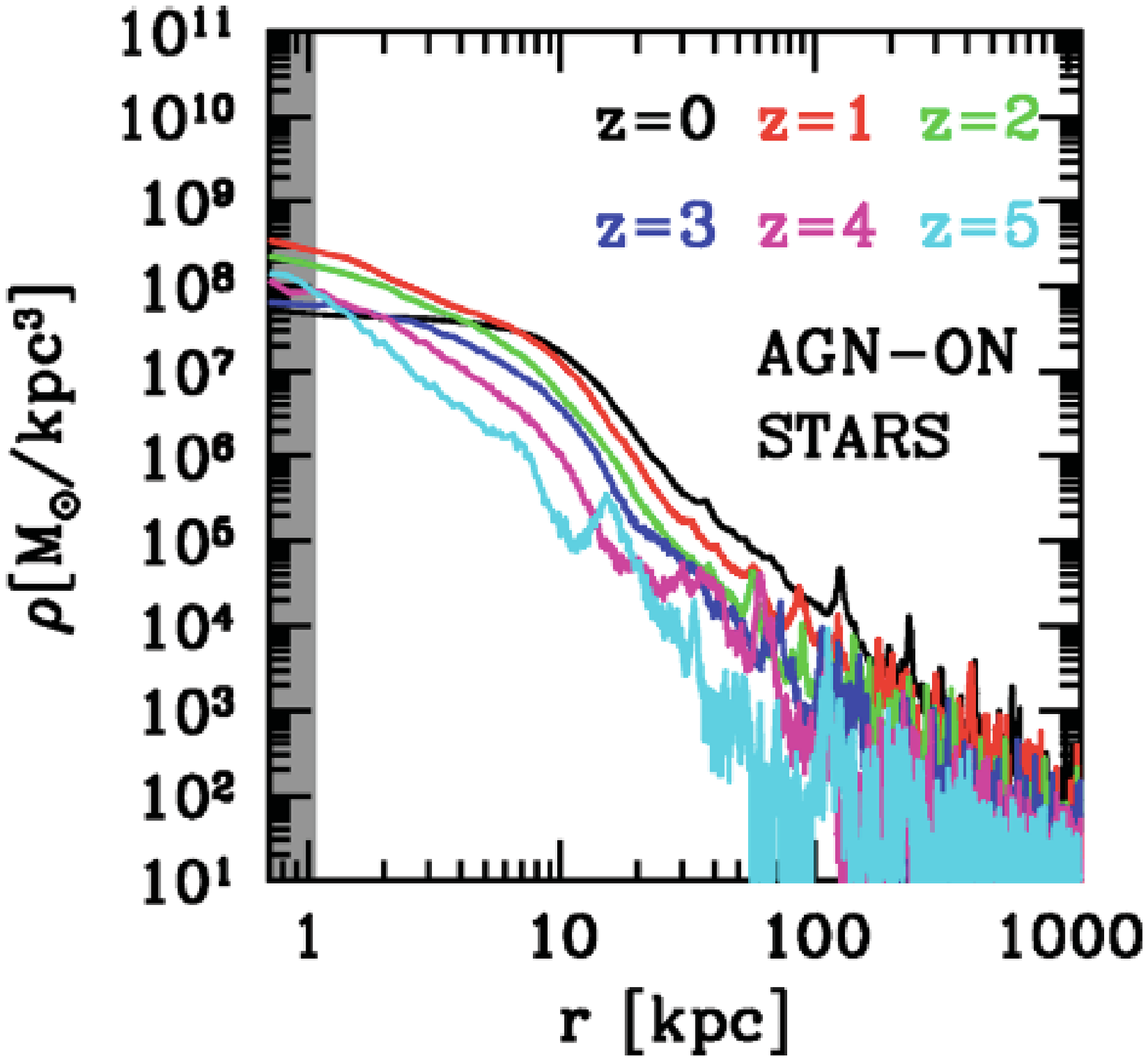} 
 \caption{{\it Left:} Size evolution driven by rapid mass-loss from
   the idealized simulations of isolated galaxies. If 40 per cent of the mass
   is lost ($\epsilon = 0.6$) the sizes can rapidly increase by 60 per
   cent. From the black to the green line the ejection varies from
   immediate ejection to an ejection time of 80 Myrs (taken from
   \citealp{2011MNRAS.414.3690R}). {\it Right:} Evolution of the stellar
   surface density profiles of a cosmological zoom-simulation of a
   brightest cluster galaxy in 
   a model with strong AGN feedback. Due to the gas explusion from the
   AGN the system is significantly more extended than in the no-AGN
   case and even develops a central core (taken from
   \citealp{2012MNRAS.422.3081M}).  
   \label{fig2}}
\end{center}
\end{figure}

\begin{figure}[t]
\begin{center}
 \includegraphics[width=2.8in]{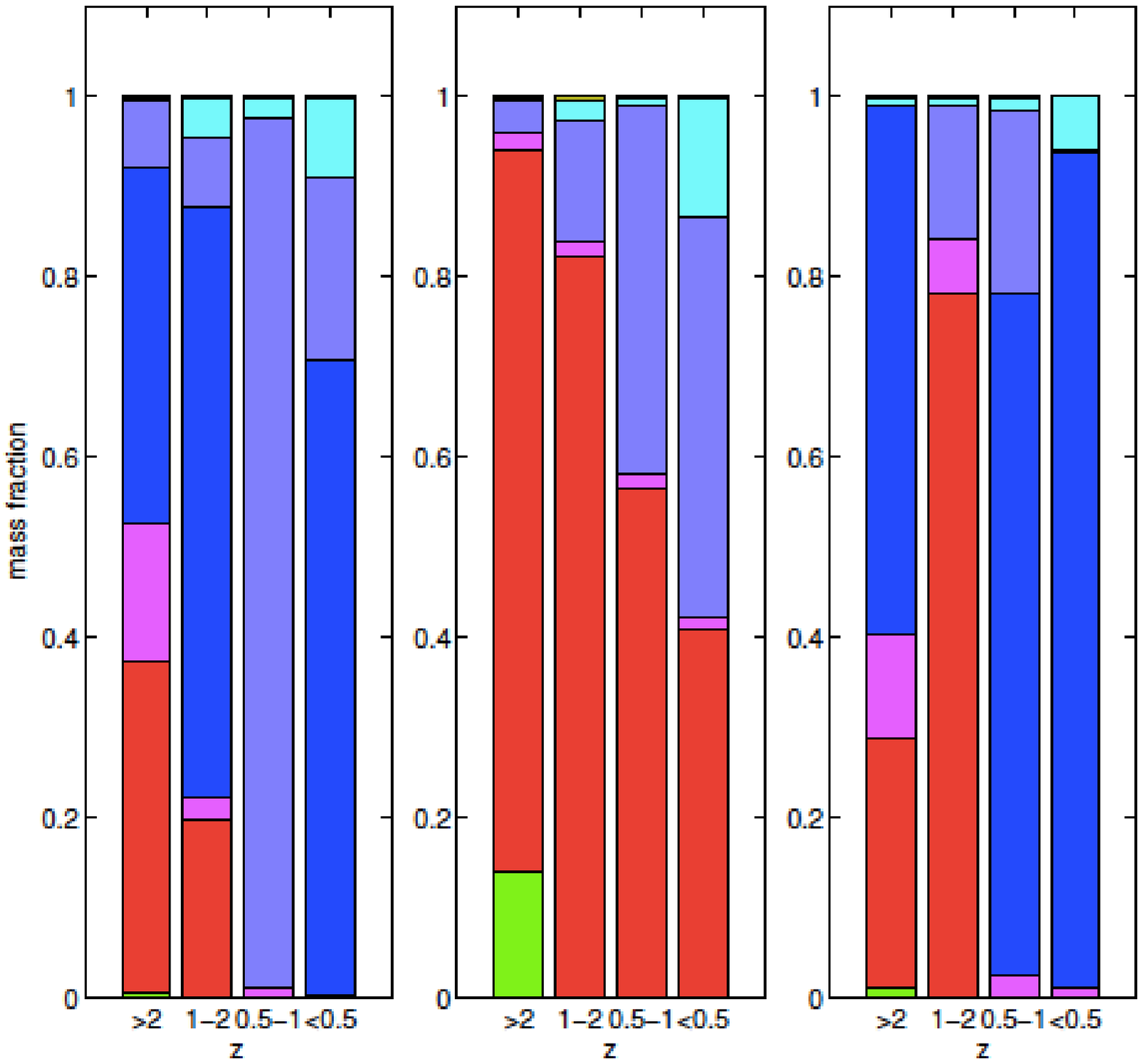} 
 \includegraphics[width=2.0in]{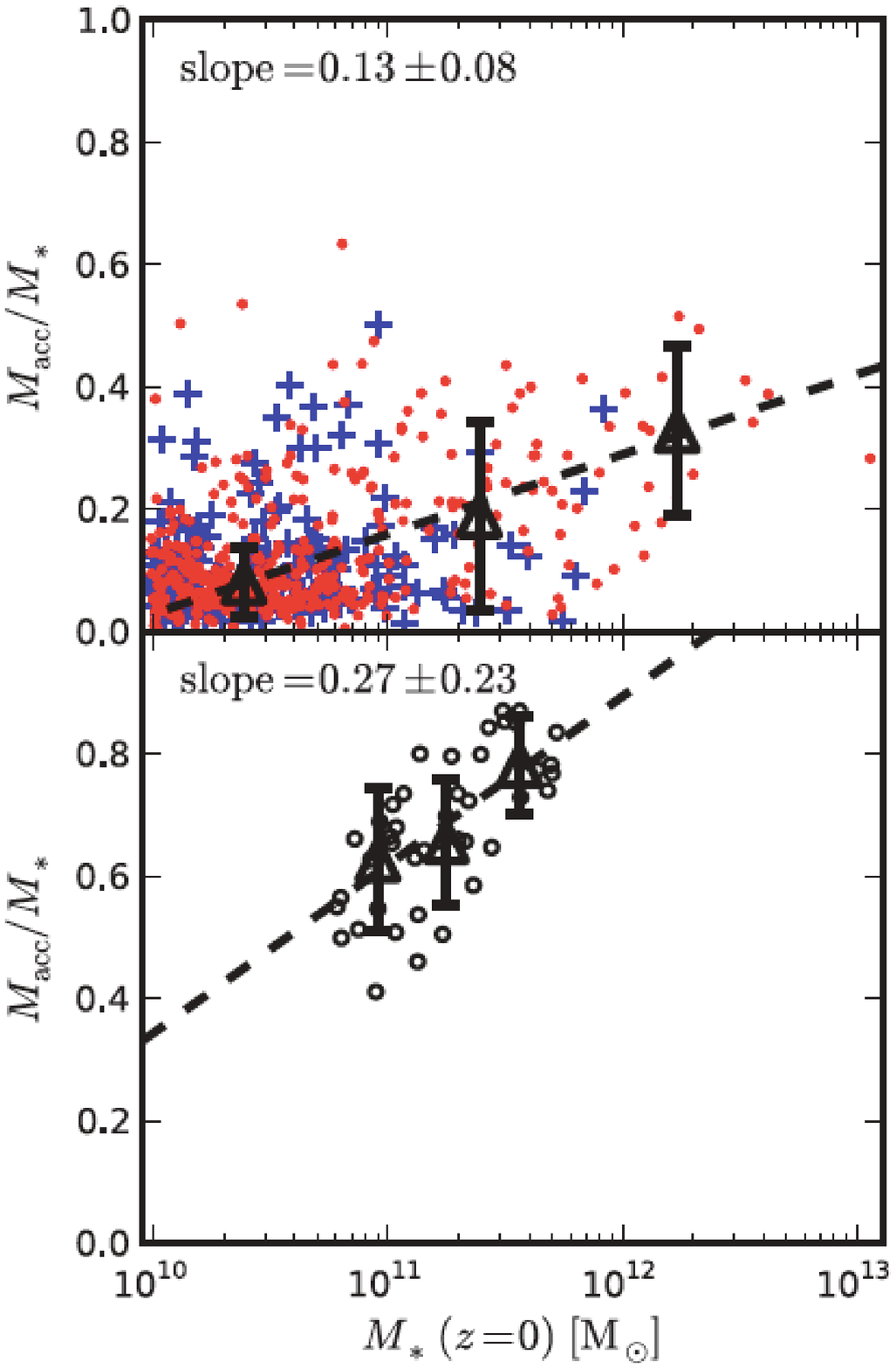} 
 \includegraphics[width=2.2in]{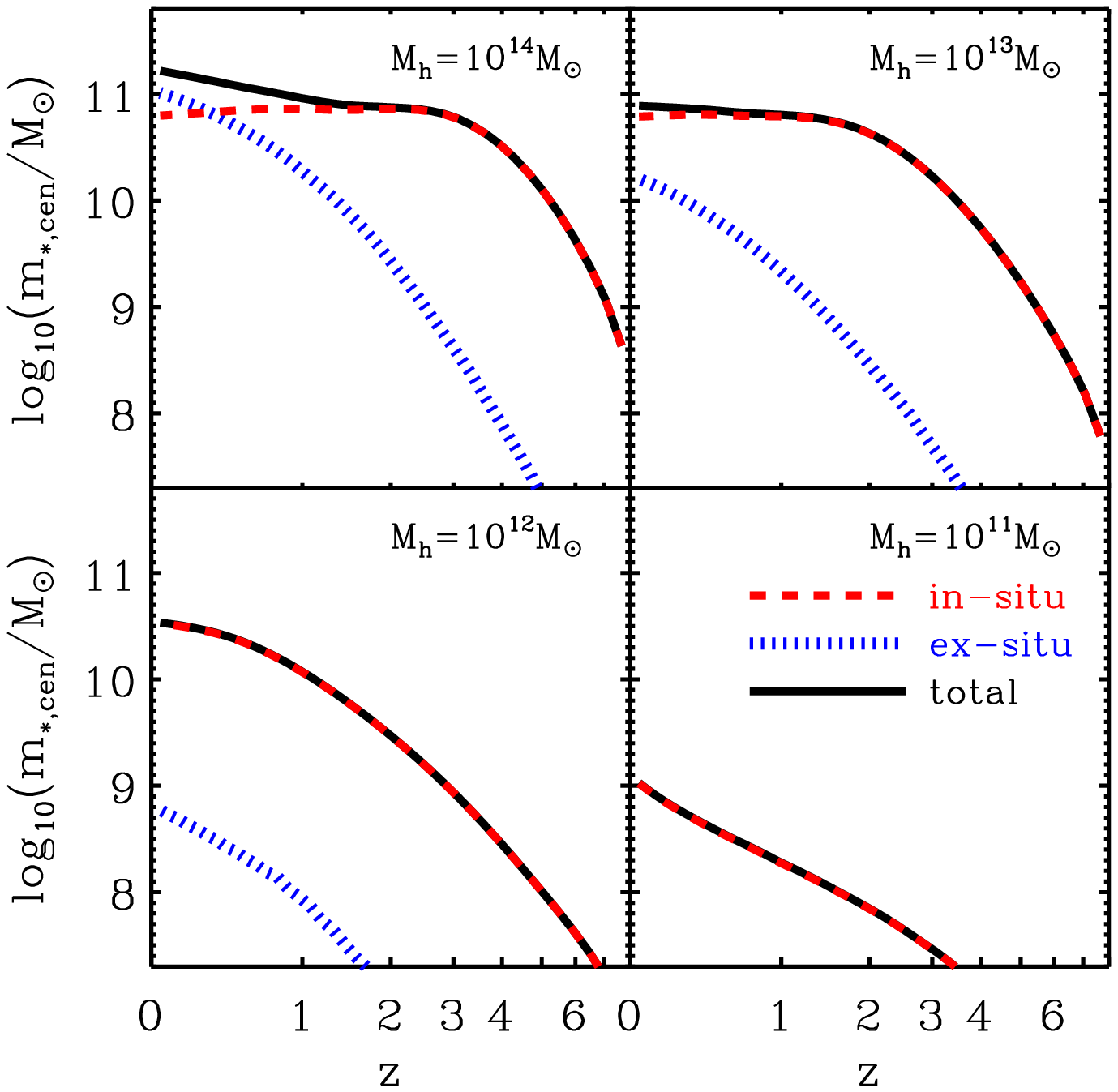} 
 \includegraphics[width=2.6in]{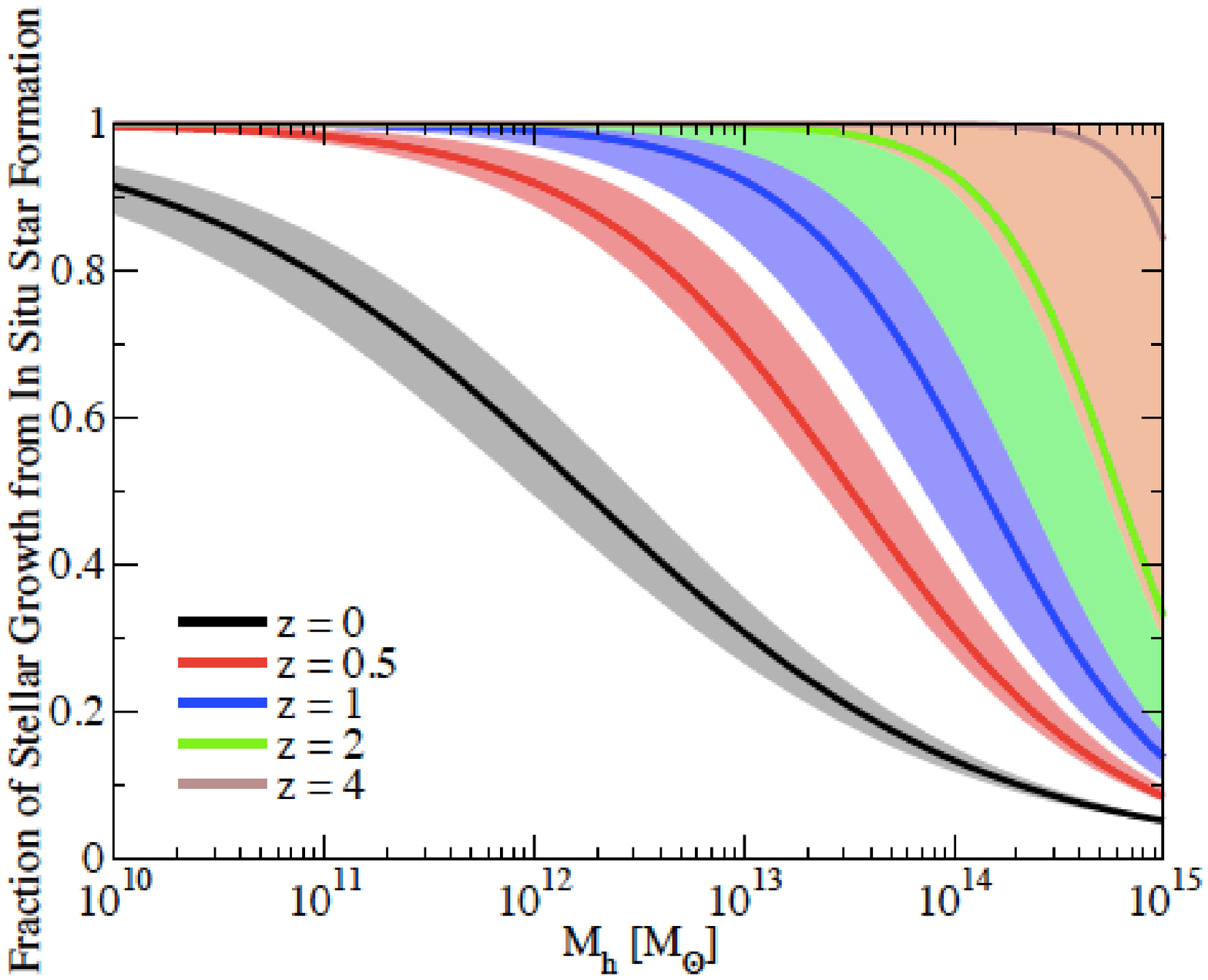} 
 \caption{{\it Upper left:} Examples for the assembly history (stellar
   origin) of three massive galaxies in high-resolution cosmological
   zoom simulations. At high redshift the formation is dominated by
   in-situ star formation (red colors). The low redshift assembly is
   dominated by merging of stellar systems  (blue colors, taken from
   \citealp{2010ApJ...709..218F}). {\it Upper right:} Ratio of the accreted
   over final stellar mass versus final stellar mass for galaxies in a
   cosmological simulation box (void: blue, cluster: red) including
   strong supernova feedback (upper panel). The fraction of accreted
   stars is about a factor 2 -3 lower than in the high-resolution zoom
   simulations of \citet{2010ApJ...725.2312O} without strong supernova
   feedback (lower panel); the trend with mass is similar but
   less strong (taken from \citealp{2012MNRAS.425..641L}). {\it Lower
     left:} Independent estimate of the ratio of accreted to in-situ
   formed stars as a function of halo mass from abundance matching
   studies \citep{2012arXiv1205.5807M}. {\it Lower right:} Similar
   estimates from a study by \citet{2012arXiv1207.6105B}. Both studies
   find a strong trend that the assembly of galaxies in more massive
   halos is more dominated by the accretion of stars rather than
   in-situ star formation. 
   \label{fig3}}
\end{center}
\end{figure}

A potential candidate for such an additional process is AGN driven
outflow of gas from a massive high-redshift gas-rich and compact galaxy
\citep{2008ApJ...689L.101F,2010MNRAS.401.1099H,2010ApJ...718.1460F}. 
In general, stellar systems suffering from central mass-loss
$\epsilon_{\mathrm{loss}} = M_{\mathrm{final}}/ M_{\mathrm{initial}}$
will expand \citep{1980ApJ...235..986H} and for rapid and slow
mass-loss simple relations for the ratio of the final to the initial
radius can be derived:   

\begin{eqnarray}
\frac{R_{final,rapid}}{R_{initial}} = \frac{\epsilon_{loss}}{2
  \epsilon_{loss} -1}, \nonumber \\
\frac{R_{final,slow}}{R_{initial}} = \frac{1}{\epsilon_{loss}}.  \nonumber
\end{eqnarray}

It is worth noting that rapid mass-loss of more than half the total
mass can unbind the whole system. This process is well known and
has been studied for star clusters \citep{1980ApJ...235..986H},
galaxies
\citep{1980ApJ...235..986H,2010MNRAS.401.1099H,2011MNRAS.414.3690R,2012MNRAS.421.3464P},
as well as cores of galaxy clusters (see Fig. \ref{fig2} and \citealp{2012MNRAS.422.3081M}).

\begin{figure}[t]
\begin{center}
 \includegraphics[width=4.9in]{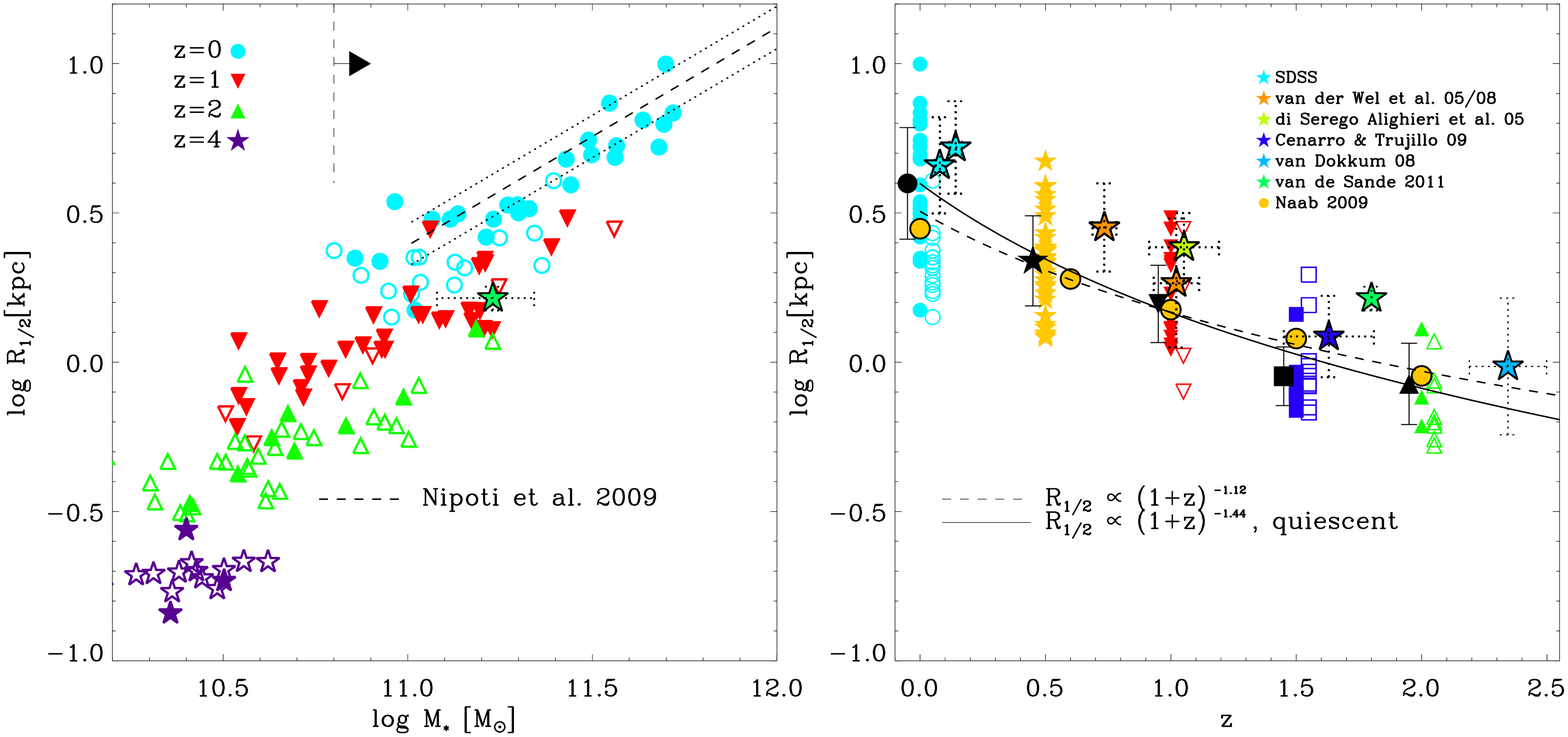} 
 \includegraphics[width=4.9in]{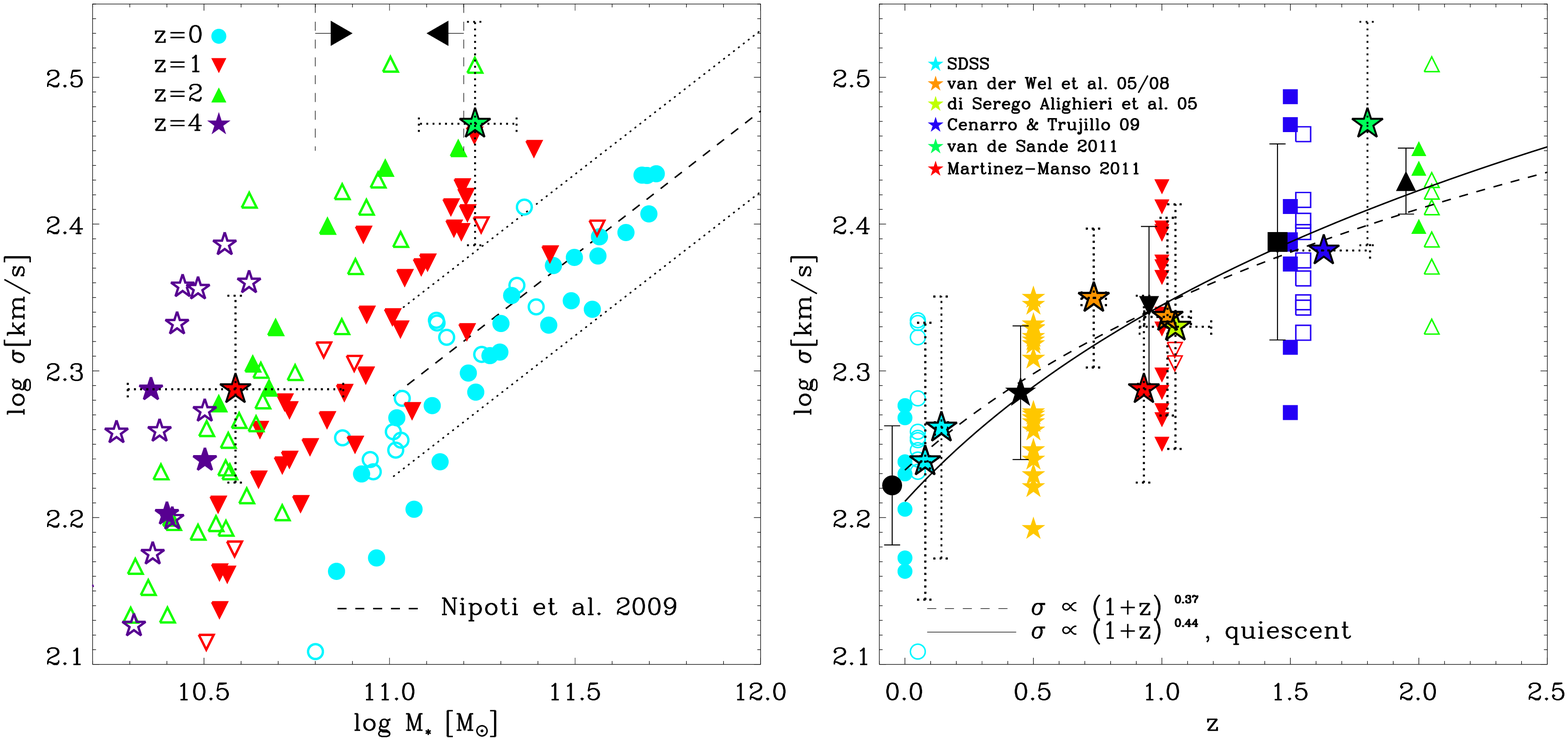} 
 \caption{{\it Upper panels:} The present day mass-size relation
   (left) for a sample of high-resolution zoom simulations (blue points, full
   symbols are quiescent galaxies) compared to observations. The evolution of the relation is driven by accretion of
   stars and is indicated by the location of the most massive
   progenitor galaxies at different redshifts.   For all galaxies more
   massive than the mass limit indicated on the left plot ($\log(M_*)
   = 10.8$) the average size evolution agrees well with observations.
{\it Lower panels:} Similar plot for the evolution of the
mass-dispersion relation (left). In a fixed mass range galaxies have
higher dispersions at higher redshifts (right). Again, the simulated
evolution is very similar to the observed one (figures are taken from
\citealp{2012ApJ...744...63O}).     
   \label{fig4}}
\end{center}
\end{figure}

\section{The cosmological two-phase assembly} 

The assembly histories of massive galaxies in currently favored
hierarchical cosmological models are significantly more complex than a
single binary merger. They grow - in particular at high redshift - by
smooth accretion of gas, major mergers but also numerous minor mergers
covering a large range of mass-ratios which can dominate the amount of
assembled stars. The picture that is emerging from from
semi-analytical models and high-resolution cosmological simulations of massive galaxies bears a
two-phase characteristic
\citep{2007MNRAS.375....2D,2008MNRAS.384....2G,2008ApJ...688..789G,2010ApJ...709..218F,2010ApJ...725.2312O,2011ApJ...736...88F,2012MNRAS.419.3200H}. 

At high redshifts the formation is dominated by dissipative processes
(i.e. significant radiative energy losses) and in-situ star formation  
leading compact progenitors with high phase space densities. In a
second phase massive galaxies are growing by the addition of stars at
large radii that have formed early outside the main galaxies in other
galaxies that were accreted later-on. This assembly phase is dominated
by collisionless dynamics and radiative energy losses are of minor
importance (see e.g. \citealp{2009ApJ...697L..38J,2010ApJ...712...88L,2012MNRAS.424..747L}). 

Independent studies using cosmological simulations based on different
numerical methods come to similar conclusions that -on average - the
mass assembly of massive galaxies is dominated by minor mergers with
mass-ratios $\sim 1:5$
\citep{2012ApJ...744...63O,2012MNRAS.425..641L,2012arXiv1202.5315G}. The 
relative importance of accreted versus in-situ formed 
stars increases with galaxy mass, a result that was already predicted
by semi-analytical models
\citep{2006MNRAS.366..499D,2007MNRAS.375....2D,2008MNRAS.384....2G}
and has been confirmed by independent estimates from abundance
matching techniques
\citep{2012arXiv1205.5807M,2012arXiv1207.6105B}. The absolute
fractions are model dependent and can vary e.g. by $\sim$ 50\% for
different feedback models
(see Fig. \ref{fig3}). Studies based on  
cosmological zoom simulations make a plausible point that the present
day scaling relations might be set by the stellar accretion history of massive
galaxies, i.e. the above mentioned fraction of in-situ to accreted
stars \citep{2012ApJ...744...63O}. In addition, based on still small
samples, high-resolution cosmological simulations the evolution of the
scaling relations appears to be in accordance with observations
\citep{2011ApJ...736...88F,2012ApJ...744...63O,2012ApJ...754..115J}. However, in general the 
cosmological simulations of massive galaxies still fail to reproduce
all observational constraints at the same time and are still limited
with respect to either resolution and statistics as well as the algorithmic
implementation of relevant feedback processes. In particular feedback
from super-massive black holes might help to finally meet observational
constraints for massive ellipticals
\citep{2010MNRAS.406..822M,2011MNRAS.412.1965M,2012arXiv1205.2694P}.    

TN acknowledges support by and valuable disucssions with Peter
Johansson, Ludwig Oser and Jeremiah P. Ostriker.

\end{document}